\documentclass[10pt,conference]{IEEEtran} 
\IEEEoverridecommandlockouts

\AtBeginDocument{%
  }

\usepackage{listings}
\usepackage{tcolorbox}
\usepackage{soul}
\usepackage{xcolor}
\usepackage{multirow}
\usepackage{fancybox}
\usepackage{xspace}
\usepackage{ragged2e}
\usepackage{changepage}
\usepackage{makecell}
\usepackage{setspace}
\usepackage{enumitem}
\usepackage[numbers,sort]{natbib}

\usepackage{booktabs, multirow} % for borders and merged ranges
\usepackage{soul}% for underlines

\usepackage{url}

\usepackage{hyperref}
\hypersetup{
    colorlinks=false,
}

\usepackage{amsmath,amssymb,amsfonts}

\usepackage{textcomp}
\usepackage{lipsum}
\usepackage{graphicx}
\usepackage{subfigure}
\usepackage{balance}
\usepackage[ruled, lined, linesnumbered, commentsnumbered, longend]{algorithm2e}
\usepackage{algorithmic}
\SetCommentSty{mycommfont}
\usepackage{array}

\usepackage{adjustbox}
\usepackage{threeparttable}

\newcommand{\logllm}{LibreLog\xspace}

\newcommand{\tbase}{LLMParser$_{T5Base}$\xspace}

\newcommand{\phead}[1]{\vspace{1mm} \noindent {\bf #1}}

\newcolumntype{?}{!{\vrule width 1.5pt}}

\newcommand{\rqboxc}[1]{\begin{tcolorbox}[left=1pt,right=1pt,top=1pt,bottom=1pt,colback=gray!5,colframe=gray!40!black,before skip=5pt,after skip=0pt]#1\end{tcolorbox}}

\definecolor{lightgray}{gray}{0.9}

\lstdefinelanguage{XML}
{
  morestring=[b]",
  morestring=[s]{>}{<},
  morecomment=[s]{<?}{?>},
  stringstyle=\color{black},
  identifierstyle=\color{darkblue},
  keywordstyle=\color{cyan},
  morekeywords={xmlns,xsi,schemaLocation}% list your attributes here
}

\begin{document}

%\title{Unsupervised Log Parsing with Open-Source LLMs for Enhanced Accuracy and Privacy}
%\author{petertsehsun }
%\date{March 2024}

\title{LibreLog: Accurate and Efficient Unsupervised Log Parsing Using Open-Source Large Language Models}

\author{\IEEEauthorblockN{Zeyang Ma}
\IEEEauthorblockA{Software PErformance, Analysis \\
and Reliability (SPEAR) Lab\\
Concordia University\\
Montreal, Quebec, Canada\\
m\_zeyang@encs.concordia.ca}
\and
\IEEEauthorblockN{Dong Jae Kim}
\IEEEauthorblockA{DePaul University\\
Chicago, Illinois, USA\\
djaekim086@gmail.com}
\and
\IEEEauthorblockN{Tse-Hsun (Peter) Chen}
\IEEEauthorblockA{Software PErformance, Analysis \\
and Reliability (SPEAR) Lab\\
Concordia University\\
Montreal, Quebec, Canada\\
peterc@encs.concordia.ca}}

\maketitle
\pagestyle{plain}

%\acmISBN{978-1-4503-XXXX-X/18/06}

\begin{abstract}

Log parsing is a critical step that transforms unstructured log data into structured formats, facilitating subsequent log-based analysis. 
Traditional syntax-based log parsers are efficient and effective, but they often experience decreased accuracy when processing logs that deviate from the predefined rules.
Recently, large language models (LLM) based log parsers have shown superior parsing accuracy. 
However, existing LLM-based parsers face three main challenges: 1) time-consuming and labor-intensive manual labeling for fine-tuning or in-context learning, 2) increased parsing costs due to the vast volume of log data and limited context size of LLMs, and 3) privacy risks from using commercial models like ChatGPT with sensitive log information.
%However, existing LLM-based parsers often require sampling and manual labeling logs for fine-tuning or in-context learning, which is time-consuming and labor-intensive. The sheer volume of log data and the limited context size of LLMs lead to increased parsing costs, both in terms of time and money, as token consumption grows linearly with log size. Moreover, most LLM-based parsers use commercial models like ChatGPT, posing privacy risks due to the sensitive information in logs.
To overcome these limitations, this paper introduces \logllm, an unsupervised log parsing approach that leverages open-source LLMs (i.e., Llama3-8B) to enhance privacy and reduce operational costs while achieving state-of-the-art parsing accuracy. \logllm first groups logs with similar static text but varying dynamic variables using a fixed-depth grouping tree. It then parses logs within these groups using three components: i) similarity scoring-based retrieval augmented generation: selects diverse logs within each group based on Jaccard similarity, helping the LLM distinguish between static text and dynamic variables; ii) self-reflection: iteratively query LLMs to refine log templates to improve parsing accuracy; and iii) log template memory: stores parsed templates to reduce LLM queries for improved parsing efficiency. Our evaluation on LogHub-2.0 shows that \logllm achieves 25\% higher parsing accuracy and processes logs 2.7 times faster compared to state-of-the-art LLM-based parsers. In short, \logllm addresses privacy and cost concerns of using commercial LLMs while achieving state-of-the-arts parsing efficiency and accuracy.

\end{abstract}

\section{Introduction}
\label{sec:introduction}

% \peter{I moved the paragraphs from related work to here. I think these are very well written and can be a good motivation in the intro.}

\begin{figure}
\centering
\includegraphics[width=\columnwidth]{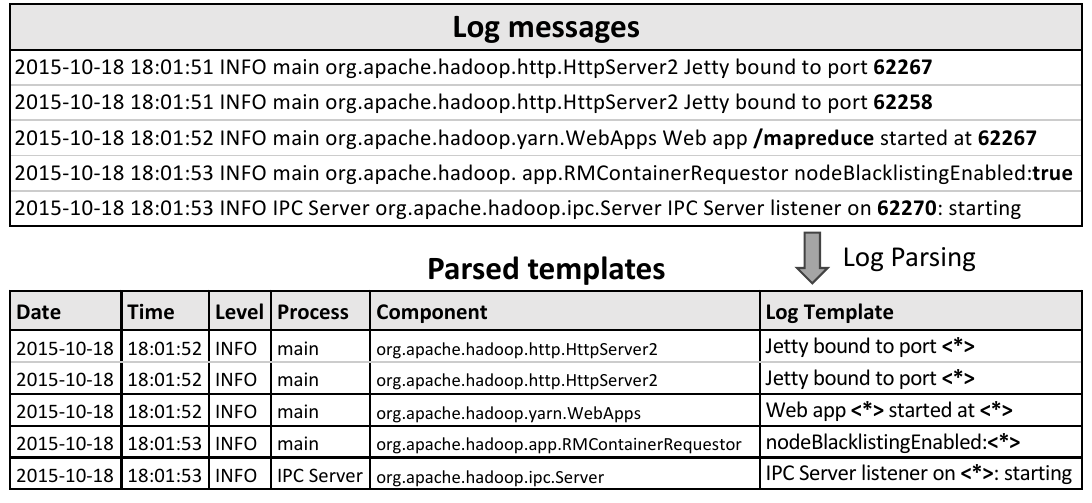}
\caption{An example of log parsing result from Hadoop.
% \peter{we need to change the logs and the arrow type, otherwise they are the same}
}
\label{fig1:definition}
\vspace{-0.2cm}
\end{figure}

Real-world software systems generate large amounts of logs, often hundreds of gigabytes or even terabytes per day~\cite{loghub2, logram, tools}. These logs provide developers with invaluable runtime information, essential for understanding system execution and debugging. To manage and analyze this vast amount of data, researchers and practitioners have proposed many automated approaches, such as monitoring~\cite{chen2019experience, wang2021would}, anomaly detection~\cite{lee2021lanobert,su2024large}, and root cause analysis~\cite{yuan2010sherlog, roy2024exploring}. 
However, as shown in Figure~\ref{fig1:definition}, logs are semi-structured, containing a mixture of static text and dynamically generated variables (e.g., port number {\sf 62267}), which makes direct analysis challenging. 

Log parsing is a critical first step in log analysis that transforms unstructured logs into log templates, dividing logs into static parts (static messages) and dynamic parts (variables). As illustrated in Figure~\ref{fig1:definition}, log templates represent the event structure of logs, providing a standardized format that simplifies further analysis. By distinguishing between static and dynamic components, log parsing enables more efficient and accurate downstream tasks~\cite{khan2023impact,10.1007/978-3-030-88494-9_16,10.1109/ICSE48619.2023.00078}. 
Given the sheer volume and diversity of generated logs, prior research has proposed various syntax-based parsers for efficient and effective log parsing. These parsers, such as Drain~\cite{Drain} and AEL~\cite{AEL}, use manually crafted heuristics or predefined rules to identify and extract log templates. Although promising, these log parsers often experience decreased accuracy when processing logs that deviate from predefined rules~\cite{tools,guidelines,loghub2}.

Recent advances in large language models (LLMs) have enabled researchers to leverage these models for log parsing~\cite{lilac,LLMParser,logppt,divlog,chatgpt_howfar,Hooglle}. %Built on transformer architectures~\cite{llama3, gpt3, codellama}, 
LLMs exhibit superior capabilities in understanding and generating text, making them particularly effective for parsing semi-structured log data. Consequently, LLM-based log parsers often achieve higher accuracy than traditional syntax-based parsers~\cite{lilac,LLMParser,logppt}. However, the sheer volume of log data and the limited context size of LLMs lead to increased parsing costs, both in terms of time and money, as token consumption grows linearly with log size. This makes practical adoption challenging. Additionally, these parsers frequently require manually derived log template pairs for in-context learning, adding significant manual overhead. 

A further complication arises from the reliance on commercial LLMs like ChatGPT by many LLM-based log parsers~\cite{lilac, divlog, chatgpt_howfar}. While powerful, using commercial models poses potential privacy risks, as logs often contain sensitive information about the software's runtime behavior and data. Uploading logs and other sensitive information (e.g., code for refactoring and bug fixes) to commercial LLMs can expose a company's sensitive data to potential privacy breaches~\cite{SamsungB90:online}.

%Therefore, employing open-source LLMs is advisable to mitigate such privacy concerns. 

To address these challenges, we propose an unsupervised log parsing technique, \logllm, which does not need any manual labels. %It uses open-source LLMs to enhance data privacy and reduce monetary costs while maintaining both parsing accuracy and efficiency. 
\logllm leverages smaller-size open-source LLMs (e.g., Llama3-8B~\cite{llama3}) to enhance privacy and reduce operational costs while achieving state-of-the-art parsing accuracy and efficiency. 
Inspired by the effective grouping capabilities of syntax-based unsupervised log parsing methods~\cite{Drain}, \logllm first groups logs that share syntactic similarity in the static text, but vary in the dynamic variable, using a fixed-depth grouping tree. %The grouping step aims to reduce parsing complexity and improve efficiency.
%\logllm leverages a smaller-size open-source LLM (i.e., Llama3-8B~\cite{llama3}) to enhance privacy and reduce operational costs. 
Then, \logllm parses logs within individual groups through three key steps: (i) \logllm uses similarity scoring-based \textit{\textbf{retrieval augmented generation (RAG)}} to select the most diverse logs based on Jaccard similarity within each log group. This step helps LLMs separate dynamic and static text by highlighting variability in dynamic variables among logs in the same group. 
%(ii) \peter{needs to be updated}it uses \textit{\textbf{in-context}} learning by providing a list of handcrafted example logs as input and corresponding log templates as output to standardize LLM's response, and
(ii) \logllm uses \textit{\textbf{self-reflection}}~\cite{shinn2024reflexion} to improve LLM responses, thereby improving parsing results. iii) \logllm uses \textit{\textbf{log template memory}} to store parsed log templates. %, thereby improving parsing efficiency. 
This approach allows logs to be parsed by first matching them with stored templates, minimizing the number of LLM queries and significantly enhancing parsing efficiency.

The paper makes the following contributions: 

\begin{itemize}
    \item We introduce, \logllm, an unsupervised log parsing technique that effectively addresses the limitations of existing LLM-based and syntax-based parsers. 
    \item \logllm employs open-source LLMs, specifically Llama3-8B, to enhance data privacy and reduce operational costs associated with commercial models. 
    \item Through extensive evaluations on over 50 million logs from LogHub2.0~\cite{loghub2}, %(0.8720-0.8289)/0.8289
    \logllm demonstrated a 25\% or higher parsing accuracy compared to state-of-the-art LLM-based log parsers (i.e., LILAC~\cite{lilac} and LLMParser~\cite{LLMParser}). Moreover, it is 2.75 to 40 times faster, showcasing its superior efficiency and effectiveness.
    \item \logllm's self-reflection mechanism helps improve parsing accuracy by over 7\%, showcasing the effectiveness of our prompting technique. 
    \item Our experiment using four small-size LLMs shows that Llama3-8B achieves the best overall result, highlighting its potential in log analysis. 
    %\item We conducted a parameter sensitivity analysis shows that \logllm 

\end{itemize}

In short, the paper provides a novel unsupervised log parsing approach that is both efficient and effective while ensuring data privacy and reducing operational costs. 

\phead{Paper Organization.} Section~\ref{sec:background} discusses background and related work. Section~\ref{sec:Approach} provides the design details of \logllm. Section~\ref{Sec:setup} outlines evaluation setup. Section~\ref{sec:evaluation} presents evaluation results. Section~\ref{sec:threat} discusses threats to validity. Section~\ref{sec:conclusion} concludes the paper. 

\phead{Data Availability:} We made our source code and experimental results
publicly available at: \url{https://github.com/zeyang919/LibreLog}

%In extensive evaluations on a large-scale public log dataset, OurParser demonstrated a 15\% higher parsing accuracy and required only half the parsing time compared to the state-of-the-art LLM-based log parsers. By addressing both privacy concerns and efficiency limitations, OurParser offers a robust solution for the challenges associated with LLM-based log parsing.

\section{Background and Related Work}
\label{sec:background}
In this section, we discuss the background of LLM and its privacy concerns. We then discuss related log parsing research.
%\peter{The background on log parsing can be shorter (we should organize the prior works in related work). The related work should be organized as: 1) Log Parsing 2) LLM-based Log Parsing}
\subsection{Background}

\phead{Large Language Models.}
Large Language Models (LLMs), primarily built on the transformer architecture~\cite{llama3, gpt3, codellama}, have significantly advanced the field of natural language processing (NLP). These LLMs, such as the widely recognized GPT-3 model with its 175 billion parameters~\cite{gpt3}, are trained on diverse text data from various sources, including source code. The training involves self-supervised learning objectives that enable these models to develop a deep understanding of language and generate text that is contextually relevant and semantically coherent. LLMs have shown substantial capability in tasks that involve complex language comprehension and generation, such as code recognition and generation~\cite{lin2024llm, abedu2024llm}. 
% \peter{a few sentences to say it is used for log analysis tasks because logs are semi-structured text, and log parsing is one of the main areas because of its importance for downstream log analysis (and cite)}
Due to logs being semi-structured texts composed of natural language and code elements, researchers have adopted LLMs to tackle log analysis tasks, such as anomaly detection~\cite{liu2023scalable,lee2021lanobert,su2024large}, root cause analysis~\cite{sarda2023adarma,sarda2023leveraging,roy2024exploring}, and log parsing~\cite{lilac,LLMParser,logppt,divlog,chatgpt_howfar,Hooglle}. Log parsing is one of the primary tasks of focus in this area, given its crucial role for more accurate and insightful downstream log analysis~\cite{khan2023impact,10.1007/978-3-030-88494-9_16}.

\phead{Privacy Issues Related to LLM.} While LLMs demonstrate remarkable capabilities in processing and generating natural language and code, their application on sensitive data such as logs presents notable privacy risks, particularly with commercial models such as ChatGPT~\cite{gpt3, Security9:online}. One major concern is that data transmitted to these models--such as system logs--could be retained and used in the model’s further training cycles without explicit consent or knowledge of the data owners~\cite{aicardi2020trust}. More importantly, sensitive data uploaded to the LLM providers could potentially be exposed through inadvertent data leaks or malicious attacks~\cite{10.1145/3660818}, posing significant privacy risks. To avoid such risks, an industry norm is to restrict the use of commercial LLMs despite their advanced capabilities. For example, Samsung bans ChatGPT and other commercial chatbots after a sensitive code leak~\cite{SamsungB90:online}. Major financial institutions like Citigroup and Goldman Sachs have restricted the use of ChatGPT due to concerns over data privacy and security~\cite{WallStre81:online}. %This practice could lead to unintended data breaches and misuse of sensitive information.
In contrast, open-source LLMs, such as those developed by Meta's Llama series~\cite{llama3,codellama}, offer greater privacy and security. Users can adopt the LLMs for local deployment to ensure data privacy, aligning with stringent data protection standards. 
%For enterprises dealing with sensitive information, this transparency is crucial to maintaining control over data and complying with privacy regulations.
Thus, open-source LLMs are more secure and trustworthy for handling confidential data such as logs\cite{Yao2023ASO,mo-etal-2024-trustworthy}. 

\subsection{Related Work}Current automated log parsers can be broadly categorized into two types: syntax-based log parsers and semantic-based log parsers. Syntax-based log parsers~\cite{AEL,Drain,Spell,logram} typically employ heuristic rules or conduct comparisons among logs to identify common components that serve as templates. Semantic-based log parsers~\cite{LLMParser,logppt,lilac,liu2022uniparser} focus on analyzing the textual content within logs to distinguish between static and dynamic segments (i.e., using LLMs), thereby deriving the log templates. Semantic-based parsers often require a data-driven approach to better grasp the semantic nuances inherent in the specific system logs they analyze.
Below, we discuss related work and the limitations of these two groups of parsers. %in two directions: syntax-based Log parsers and semantic-based log parsers.

\phead{Syntax-based Log parsing approaches.}
Syntax-based log parsers~\cite{AEL,Drain,logram,Spell} generally utilize manually crafted heuristics or compare syntactic features between logs to extract log templates. 
Different from general text data, log messages have some unique characteristics. Heuristic-based log parsers extract log templates by identifying features in the logs.
For example, AEL~\cite{AEL} uses heuristics to remove potential dynamic variables and extract log templates. Drain~\cite{Drain} employs a fixed-depth parsing tree structure alongside specifically designed parsing rules (i.e., top-k prefix tokens) to identify common templates. 
However, these log parsers often suffer from decreased accuracy when processing logs that do not conform to the predefined rules. 

Logs with the same log template share the same static messages in the log. Based on this observation, several log parsers leverage frequent pattern mining~\cite{Spell, logram} to parse the logs by identifying common textual content within logs. 
For instance, Spell~\cite{Spell} uses the Longest Common Subsequence to parse logs, and Logram~\cite{logram} identifies frequent $n-gram$ patterns within logs, using these recurring patterns to parse logs. While these frequent pattern mining-based parsers do not require manually defined rules, the templates they generate are highly dependent on the structure of the input logs. Logs with complex structures may lead to poor frequent pattern mining results, resulting in low parsing accuracy. 
\textit{\textbf{In short, while syntax-based parsers benefit from simplicity and efficiency in identifying common templates, their performance varies depending on the structure of logs.
}}
\phead{Semantic-based log parsing approaches.}
%\peter{we should cite more, and maybe shorten the discussion for each work a bit}\peter{Log parsing: How far can chatgpt go?}\peter{DivLog: Log Parsing with Prompt Enhanced In-Context Learning}\peter{High-precision Online Log Parsing with Large Language Models}
Semantic-based log parsers~\cite{lilac,LLMParser,logppt,divlog,chatgpt_howfar,Hooglle} use language models to analyze the semantics of the log messages for log parsing. %\peter{what they are, like leverage ...}. 
Recently, they have shown superior parsing accuracy compared to syntax-based log parsers, largely due to significant advancements in language models. 
%Large Language Models (LLMs) have demonstrated considerable success in log parsing tasks by leveraging their advanced semantic processing capabilities~\cite{lilac,LLMParser,logppt,divlog,chatgpt_howfar,Hooglle}. 
For instance, models like ChatGPT~\cite{chatgpt_howfar} can analyze the context of log messages and dynamically generate log templates without prior knowledge, enhancing accuracy and adaptability across different log formats. %Applying LLMs in log parsing facilitates a deeper understanding of unstructured log data, which is crucial for effective log management and anomaly detection in complex systems.
%Le and Zhang~\cite{chatgpt_howfar} further evaluated using ChatGPT to parse logs. 
DivLog~\cite{divlog} enhances log parsing by extracting similar logs from a candidate set of labeled logs for in-context learning using GPT-3~\cite{gpt3}. 
Due to the high cost of commercial LLMs such as  ChatGPT, LILAC~\cite{lilac} enhances the efficiency of LLM-based log parsing by incorporating an Adaptive Parsing Cache that stores parsing results. LILAC adopts in-context learning with log-parsing demonstrations (i.e., manually created log templates) for enhanced parsing accuracy. 

% \peter{@Zeyang, please try to improve the paragraphs below. There seems to be repetition. }However, since logs often contain sensitive data, they cannot be shared with commercial LLMs like ChatGPT in practice. Moreover, due to the large volume of log data, the monetary cost of using ChatGPT can be significant. Hence, some parsers aim to use open-source LLMs. 
Some parsers also aim to use open-source LLMs for log parsing. 
Hooglle~\cite{Hooglle} adopted an LLM pre-trained on labeled logs for log parsing. LogPPT~\cite{logppt} utilizes a masked language model (RoBERTa~\cite{Roberta}) and adopts few-shot learning to classify tokens in log messages based on few-shot examples. As an initial attempt to apply LLMs for log parsing, LogPPT showed improved accuracy over traditional syntax-based log parsers. 
LLMParser~\cite{LLMParser} explores the performance of various LLMs after a few-shot fine-tuning on log parsing. Results indicate that fine-tuning small open-source LLMs with a few demonstrations can also achieve high log parsing accuracy.

\textbf{\textit{Although the results are promising, recent works in semantic-based log parsers have two main limitations: 1) privacy and monetary costs of using commercial LLMs and 2) requiring manually derived log templates for LLMs to learn}}. First, most log parsers are based on commercial LLMs such as ChatGPT, which makes real-world adoption a challenge due to the privacy issues and monetary costs of parsing large volumes of logs. 
Second, many parsers, especially the ones that aim to improve efficiency and accuracy (e.g., LILAC~\cite{lilac}) or the ones that use smaller open-source models (e.g., LogPPT~\cite{logppt} and LLMParser~\cite{LLMParser}) require some log-template pairs as the demonstration. Deriving such templates requires significant manual efforts, and the provided demonstrations may affect the parser's accuracy on logs with unseen templates. 

In this paper, we propose \logllm that addresses the two above-mentioned limitations. We deployed a relatively small open-source LLM (i.e., Llama3-8B~\cite{llama3}) on log parsing to avoid privacy issues and monetary costs. Additionally, \logllm enhances LLM-based log parsing by capitalizing on the commonalities and variabilities within logs to provide a demonstration-free prompt for the LLM.

%parsing accuracy may be biased towards 

%Real-world software systems generate large amounts of logs, often hundreds of thousands per minute~\cite{loghub2,logram,tools}. It is crucial for log parsers to maintain high efficiency when handling such large volumes of data. Although LLM-based log parsers show promising results, the activities involved, such as log labelling (for demonstrations) and frequent model queries, significantly increase time and economic costs. 
% time, economic and resource consumption. 
% With the help of a CO2 emissions calculator~\cite{MachineL18:online}, it is estimated that each message sent to ChatGPT produces approximately 4.32 grams of CO2~\cite{Examinin20:online}. The carbon footprint of 100 ChatGPT queries is comparable to the emissions generated by driving a car for one mile~\cite{Greenhou14:online}. 

% Thus, several crucial challenges for LLM-based log parsers are as follows: (1) reducing the frequency of LLM queries and (2) diminishing reliance on labeled logs to reduce burdens for domain experts. Addressing these challenges is essential to enhance the practicality of LLM-based log parsing approaches.
%\rqboxc{LLM-based log parsers require manually labeled logs for supervised enhancing the accuracy. Furthermore, adopting LLMs for parsing large volumes of logs poses efficiency and economic  impacts due to the computational demands of these models. Additionally, the use of closed-source LLMs for log parsing raises privacy concerns, as sensitive log data might be leaked and stored by model providers.}

\section{Approach}~\label{sec:Approach}

% The core concept of \logllm involves the grouping similar logs into distinct log groups. Subsequently, similar logs are sent to an LLM for parsing, which generates a log template by analyzing the commonalities and differences among the grouped logs. This template is then verified for its ability to accurately match the logs within its group. In instances where mismatches occur, the LLM undertakes a self-correction process. This involves re-sampling the mismatched logs and sending them back to the LLM to generate a revised log template~\djk{how does resampling help mitigate mismatch exactly}~\zeyang{because we remove the matched logs in the groups, so we need to re-sample the mismatching logs and send them to LLM for parsing again.}. Successfully matched log templates are incorporated into a \djk{does resampling get re-incorporated into existing group of logs or does it create new one?}~\zeyang{the initial sampling is sample logs from one group (send these samples to LLM for parsing). then we check the generated log template with all the logs in the group. if some logs match, we remove them from the group (consider them correctly parsed). If there are still some mismatched logs in the group, we re-sample them and iterate the prior steps to parse the logs utill the group is empty.} log template pool. This pool serves as a repository that subsequent log groups can query to retrieve existing templates, thereby reducing the need for repeated LLM queries and enhancing the efficiency of log parsing.

\begin{figure*}[h]
\centering 
\includegraphics[width=1\linewidth]{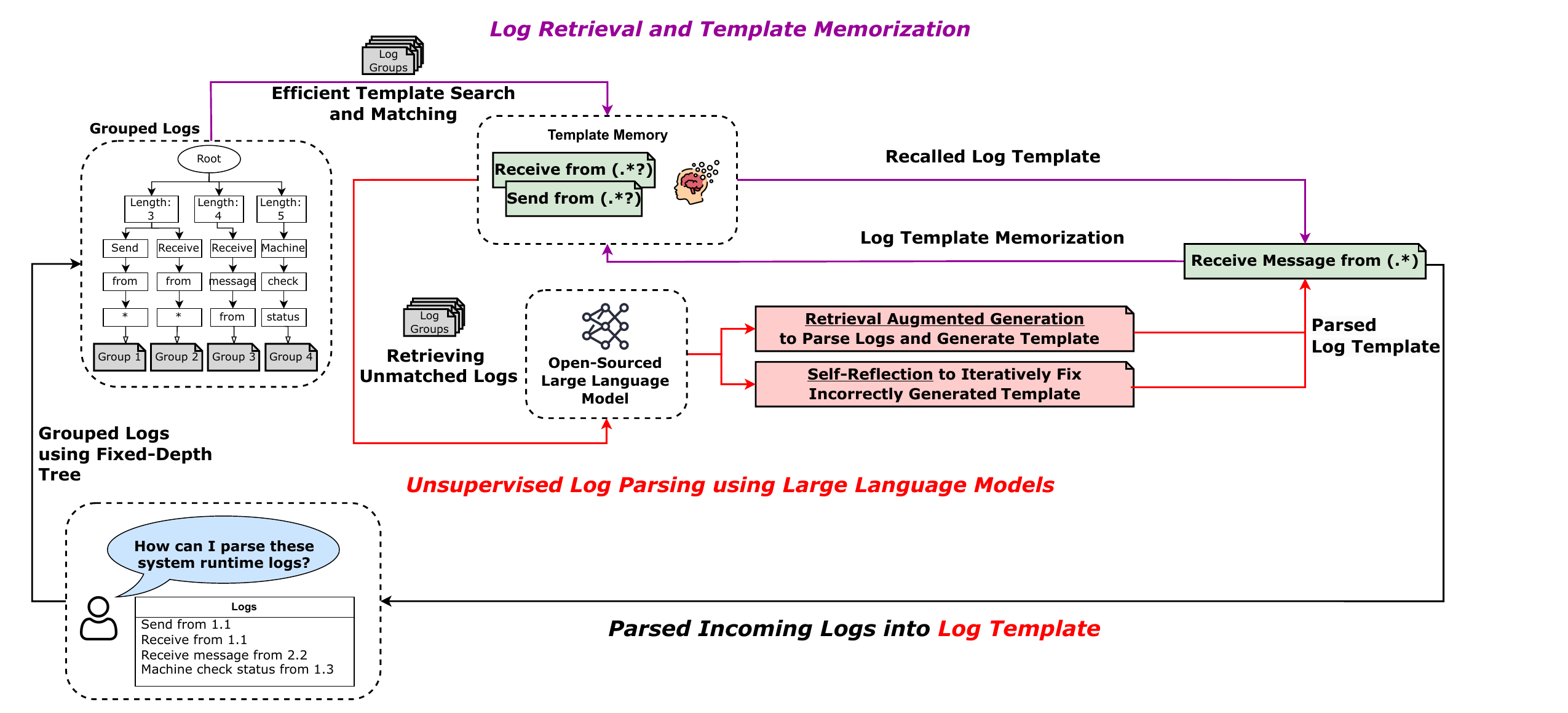} 
\caption{An overview of \logllm.}
\label{fig:overall}
\end{figure*}

In this section, we introduce \logllm, an efficient unsupervised log parser, leveraging memory capabilities and advanced prompting techniques to maximize efficiency and parsing accuracy. \logllm leverages a smaller-size open-source LLM to enhance privacy and reduce operation costs. Figure~\ref{fig:overall} illustrates the overall architecture of \logllm, which primarily comprises of three components: 
(i) \textsf{\textbf{log grouping}}, which groups logs that share a commonality in their text. Such log groups can then be used as input to LLM to uncover dynamic variables.  
(ii) An \textsf{\textbf{unsupervised LLM-based log parser}} that uses retrieval-augmented generation (RAG), followed by an iterative self-reflection mechanism to accurately parse the grouped logs into log templates. 
(iii) An \textsf{\textbf{efficient log template memory}}, which memorizes the parsed log templates for future query. The core idea is to enhance efficiency by storing parsed log templates in memory, thereby avoiding the need for repeated LLM queries. 

%We discuss the details of \logllm below. 

\subsection{Log Grouping Based on Commonality}
\label{sec:grouping}
% Unlike previous approaches~\cite{LLMParser, lilac, logppt} that rely on labeled data to learn domain-specific knowledge of the software system, including syntactic and semantic information, as few shots for in-context learning or fine-tuning, \logllm does not require any manual labels and thus is strictly zero-shot. 
\logllm achieves unsupervised and zero-shot log parsing by first applying an effective grouping strategy. % to group similar logs without needing labeled data. 
This strategy aims to group logs that share commonality in their static text, yet are different in their dynamic variables. Such log groups can then be used as input to LLMs to generate log templates by prompting LLMs to identify the dynamic variables among logs in the same group. To group the logs, we adapt the efficient unsupervised methodology proposed by Drain~\cite{Drain}, which applies a fixed-depth parsing tree and parsing rules (i.e., $K$ prefix tokens) to identify log groups. The fixed depth in our grouping tree provides a structured and predictable framework that enhances efficiency. By limiting the depth, we reduce the complexity of the tree traversal, which speeds up the grouping process.
%We employ a grouping \textit{tree of fixed depth} to facilitate the grouping of logs.~\djk{add a brief rationale on why fixed depth is chosen and how it aids efficiency.}\peter{and the nodes in the tree}
%\zeyang{The fixed depth in our grouping tree provides a structured and predictable framework that enhances efficiency. By limiting the depth, we reduce the complexity of the tree traversal, which speeds up the grouping process. }

Our fixed-depth tree implementation for grouping consists of three key steps: (i) group by \textit{\textbf{length}}, (ii) group by \textit{\textbf{K prefix tokens}}, and (iii) group by \textit{\textbf{token string similarity}}. In step (i), we first group the logs based on \textit{token length}, which partitions the logs into subsets of logs that are similar in token length. This initial grouping significantly reduces the computational complexity in the subsequent grouping phases. 
In step (ii), the grouped logs are then kept at a fixed depth which stores $K$ prefix tokens.
% \peter{what is the depth? do you do recursive grouping? what is the threshold to decide they belong to the same group?}.\peter{please just update the paper} 
Since logs are initially grouped based on \textit{token length}, truncating $K$ prefix tokens (default the first three tokens of the log) can limit the number of nodes visited during the subsequent traversal process for step (iii), significantly improving grouping efficiency. 
Prior to step (iii), it is important to note that we abstract the numerical literals in the logs with a wildcard symbol (*). This is done to prevent the issue of grouping explosion in step (iii), which can make grouping inefficient. Finally, in step (iii), we calculate the similarity between the new logs and the log groups stored in the fixed-depth tree. This step determines whether the incoming log fits into an existing group or necessitates the creation of a new log group. If a suitable group is found based on the similarity threshold, i.e., $\frac{\text{\# of common tokens}}{\text{total number of tokens}} > 0.5$, the log is inserted into existing log groups. If not, a new group is created, and the tree is dynamically updated to accommodate this new log pattern. This adaptive approach ensures that our system evolves with the incoming data, continuously optimizing both the accuracy and efficiency of the log grouping process. 

\subsection{LLM-based Unsupervised Log Parsing}\label{sec:parsing} %Our LLM-based log parsing combines retrieval-augmented technique and iterative self-reflection. 
\begin{figure}
\centering
\includegraphics[width=1\columnwidth]{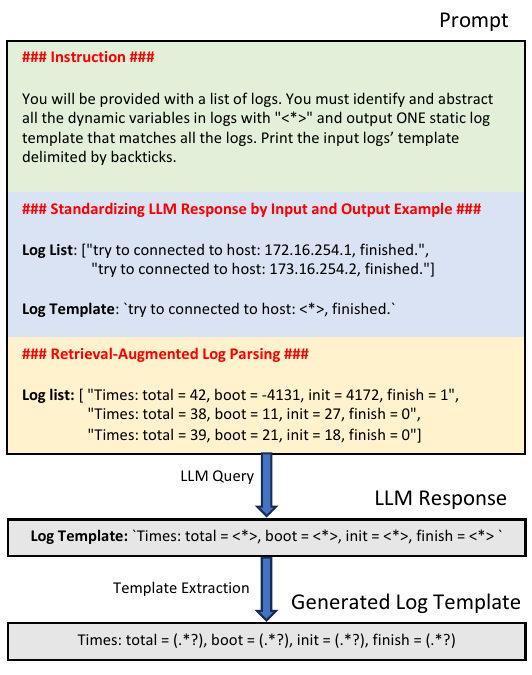}
\caption{Example of the prompt template used for \logllm. The green block illustrates the task instruction provided to the LLM. The blue block highlights the input and output examples used to standardize the log response format. The yellow block depicts the retrieval-augmented selection process that enhances log parsing accuracy by incorporating representative variability.}
\vspace{-0.1cm}
\label{fig:prompt}
\end{figure}

Our prompts to LLMs contain representative logs (based on variability) retrieved from each log group (from Section~\ref{sec:grouping}) to guide LLMs in separating dynamic variables and static text. Figure~\ref{fig:prompt} illustrates the prompt template that \logllm uses. Below, we discuss the composition of our prompt in detail.

%and uses carefully crafted prompts to guide the LLM. Unlike methods requiring fine-tuning with labeled logs, our method relies on explicit instructions and examples to enhance performance. Figure~\ref{fig:prompt} illustrates the prompt template used in our approach.

\noindent{\textit{\underline{Prompt Instruction.}}}
In the instruction part of our prompt, we define the goal of the log parsing task to the LLM (highlighted in green in Figure~\ref{fig:prompt}). We emphasize that all the provided logs should share one common template that matches all selected logs. This specification is crucial to ensure that the LLM can effectively identify the commonalities and variability within the provided logs, thereby preventing any difficulties in parsing due to inconsistent log templates.

\noindent{\textit{\underline{Standardizing LLM Response by Input and Output Example.}}}
Since our LLM is not instruction fine-tuned~\cite{LLMParser}, it is crucial to clearly describe our task instruction and include an input-output example in the prompt. This explicit guidance helps the LLM understand the desired input and output formats. 
%Existing LLM-based parsers~\cite{lilac, logppt} leverage in-context learning with system-specific manually labeled logs and corresponding templates, enhancing performance. In contrast, our method uses a generic list of handcrafted logs for all software systems and one corresponding template output example within our prompts (highlighted in blue in Fig~\ref{fig:prompt}). 
As shown in Figure~\ref{fig:prompt}, we provide one example to illustrate the input/output form. The example remains unchanged for all systems.  
This approach effectively guides the LLM in understanding the objective and input-output formats without the need of instruction fine-tuning or labeled data. %, significantly improving performance without relying on fine-tuning to system-specific labeled logs.

% \noindent{\textit{\underline{Prompt Instruction.}}}
% In the instruction part of our prompt, we provide a standardized format for both input and output, defining the goal of the log parsing task to the LLM (highlighted in green in Fig~\ref{fig:prompt}). Additionally, we emphasize that all the selected logs should share one common template that matches all selected logs. This specification is crucial to ensure that the LLM can effectively identify the commonalities and variability within the selected logs, thereby preventing any difficulties in parsing due to inconsistent log templates. 

% \noindent{\textit{\underline{Standardizing LLM Response.}}}
% To mitigate the risk of the LLM failing to fully comprehend our requirements and the specified input-output formats from the instructions alone, we enhance the prompt with a list of generic logs paired with their template as an example (highlighted in blue in Fig~\ref{fig:prompt}). Unlike previous LLM-based log parsers~\cite{lilac,logppt}, which rely on manually labelled logs from the target system to guide the model in accurately identifying log templates. Instead, all parsing processes in our approach utilize the same generic list of handcrafted logs and their template in the prompt. This method serves to guide the LLM in correctly understanding the task objectives and the formats of inputs and outputs. This design significantly improves the LLM's comprehension of the descriptions provided in the instructions.

\noindent{\textit{\underline{Retrieval-Augmented Log Parsing.}}}\label{sampling}
To parse logs accurately, we select representative logs that showcase variabilities within a log group based on commonality. By presenting the LLM with logs sharing the same structure but varying in dynamic variables, it can more effectively distinguish between fixed and dynamic elements to identify the log template. We developed a retrieval argument generation (RAG) approach based on Jaccard Similarity~\cite{tan2016introduction}. 
Jaccard similarity measures the similarity between two given sets by calculating the ratio of the number of elements (e.g., tokens) in their intersection to the number of elements in their union.
For log data, each log is split into a set of tokens (i.e., words), and then these tokens are used to determine the sizes of the intersection and union. The resulting ratio is the Jaccard similarity between two given logs, with a ratio closer to one indicating higher similarity. 
We aim to identify the logs with the greatest variability within the same group. Hence, we select logs with the lowest Jaccard similarity score. This approach helps create accurate log templates by focusing on logs that are most indicative of the entire group's characteristics.

Our selection process starts by selecting the longest log (based on the number of characters) within the group as the initial reference. We then calculate the Jaccard similarity between this log and every other log in the group. The log with the lowest similarity to the reference log is added to the selection set. We continue computing pairwise Jaccard similarity between the selected logs and the remaining unselected logs, sequentially adding the log with the lowest similarity. This iterative process is repeated until \textbf{K} (default $K=3$) logs have been selected, ensuring the selected logs effectively represent both the commonality and diversity within the log group. Given the computation costs and to ensure efficiency, we randomly select at most 200 logs from each group (or all the logs if the number is less than 200) for our log selection process. %limit the comparison to at most 200 logs. %we initially limit the group to 200 logs (i.e., randomly sample 200 logs from the log group) if it exceeds this number, ensuring efficiency while capturing representative diversity.

Specifically, the logs selected from the log group are listed in the format of a Python list within the prompt for parsing. We use a prefix (i.e., `\texttt{Log list:}') to help the LLM identify the logs that require parsing (highlighted in yellow in Figure~\ref{fig:prompt}). This consistency in input format, mirroring the ``Input and Output Example'', also guides the LLM to respond with the log template in a fixed format as demonstrated in the example, facilitating accurate template generation and extraction.

\noindent{\textit{\underline{Post-processing Template Standardization.}}}
We use a post-processing technique to further standardize the log template generated by LLM. We employ string manipulation techniques to remove non-template content from the response (i.e., prefixes and backticks). To facilitate the verification of the accuracy of log templates, we replace the placeholder "\texttt{<*>}" within the templates with the regular expression pattern "\texttt{(.*?)}". The regex template enables a direct matching process when comparing the generated templates with logs, and can be directly applied to abstract logs. 

\noindent{\textit{\underline{Self-Reflection for Verifying Log Template.}}} After generating a log template, we verify whether the template can match each log within the group. If a log is correctly matched by a log template, we consider it to be parsed successfully. The log template is then added to the log template memory for future use. After all logs in the group have been checked, any unparsed logs undergo a \textsf{self-reflection} process~\cite{shinn2024reflexion}, which aims to revise the templates and improve parsing results. Similar to the initial parsing attempt, we first select these unparsed logs and then utilize the prompt described in Figure~\ref{fig:prompt} to generate a new log template using LLMs. This step is repeated until all logs in the group can be matched/parsed by the generated templates. Note that, to prevent the LLM from entering a parsing loop (i.e., repeatedly generating incorrect templates), we limit the self-reflection process to three iterations. %This safeguard helps avoid process stagnation due to a few difficult logs. 

\subsection{Template Memory for Efficient Log Parsing}
Repeatedly using LLM to parse logs with identical groupings and templates significantly increases the frequency of LLM queries, thereby reducing the efficiency of the log parsing process. To address this issue, we introduce \textit{\textbf{log template memory}} in \logllm, which stores the parsed log templates for future parsing, avoiding redundant LLM queries.

\noindent{\textit{\underline{Efficient Log Template Memory Search and Matching.}}}
When a log group requires parsing, we first check whether a matching log template exists within the memory. If some logs within the group find a matching template in the memory, we apply this log template to parse the logs, mitigating the need for LLM queries. However, it is possible that some logs within the same group may match while others may not (e.g., due to limitations in the grouping step or limitation of the log template). Hence, the logs that remain unparsed are then sent to LLM for parsing. The new log template generated from this process is then added to the \textit{\textbf{log template memory}} for future reference. This design significantly reduces the number of LLM queries during the log parsing process.

To efficiently utilize log templates in the memory, there is a need for an efficient \textit{\textbf{search}} mechanism to verify whether or not the given logs match existing log templates in the memory. This is crucial since the memory can be large, consisting of many log templates. For every log, we need potentially at most $N$ searches for $N$ log templates. 
To improve efficiency, we put forward one key observation: the token length of log templates is always less than or equal to that of the original logs, as multiple tokens may be treated as a single variable during log parsing. For instance, consider the log \texttt{`sent 100 bytes data'}. After parsing, the corresponding log template is generated as \texttt{`sent <*> data'}. The original log consists of four tokens, whereas the parsed template has three. This reduction in token count occurs because \texttt{`100 bytes'} is treated as a single variable, thus decreasing the overall length of the template compared to the original log. Consequently, when searching for log templates in the memory, we first sort the templates based on the number of tokens. This sorting allows us to efficiently check new logs by first calculating the token length of the log to be parsed, then using binary search to find all templates with a token count less than or equal to the log length. This design reduces the number of match checks required from $O(N)$ to $O(Log N)$, thereby enhancing the efficiency of the search process. 

Our log-template matching process is efficient. Unlike traditional log templates that use placeholders (i.e., \texttt{``<*>''}) to abstract dynamic variables within logs, we store log templates in memory as regular expression patterns (i.e., use \texttt{``(.?)''} instead of placeholders). This adjustment allows us to use regular expressions to efficiently verify whether logs match with log templates in memory and improve matching efficiency. %This design improves the efficiency of the matching process.

\section{Experiment Setup}
\label{Sec:setup}
In this section, we discuss our experiment setup to answer our research questions and \logllm's implementation details. 

\phead{Studied Dataset.}
We conduct our experiment on the log parsing benchmark LogHub-2.0 provided by He et al.~\cite{he2020loghub,loghub2}. 
This benchmark contains logs from 14 open-source systems of different types, such as distributed systems, supercomputer systems, and server-side applications. LogHub is widely used to evaluate and compare the accuracy of log parsers~\cite{logram,Drain,Spell,LLMParser, lilac,logppt}. 
Compared to LogHub-1.0~\cite{he2020loghub}, the number of logs has increased significantly in LogHub-2.0, increasing from 28K (2K logs per system) to more than 50 million %50,416,620 
logs with a total of 3,488 different log templates. 
LogHub-2.0 also provides the groundtruth log template for each log. 
%At the same time, Khan et al.~\cite{guidelines} observed that the original labels in the LogHub dataset have some errors due to inconsistent labelling styles. In LogHub-2.0, these log template labelling errors are also fixed. 
With this large-scale LogHub-2.0 dataset, researchers can better evaluate the efficiency and effectiveness of log parsers~\cite{lilac, loghub2}. %the latest log parsers are~\cite{lilac} able to evaluate their effectiveness and efficiency on a more comprehensive log dataset. Therefore, our evaluation in this paper is conducted constantly on this large-scale dataset.

\phead{Environment and Implementation.}
Our experiments were conducted on an Ubuntu server with an NVIDIA Tesla A100 GPU, AMD EPYC 7763 64-core CPU, and 256GB RAM using Python 3.9. We execute the baselines using their default parameters under the same environment to compare the efficiency. We use Llama3 8B~\cite{llama3} for \logllm's underlying LLM because it is a relatively small yet powerful model, balancing performance and efficiency effectively. We set the temperature value to 0 to improve the stability of the model output.  
Note that it is easy to switch to other LLMs. In RQ4, we evaluate \logllm by replacing Llama3 with other open-source LLMs.  %Note that it is straightforward to switch to other LLMs, and we evaluate the result of using different open-source LLMs in RQ3. %\peter{mention the LLM we use and why we chose it (e.g., good at SE tasks and a relatively small size). It can also be easily replaced}%for evaluation as same as \logllm to ensure the fairness of the comparison.

\phead{Evaluation Metrics for Log Parsing. }
Following prior studies~\cite{LLMParser,lilac,logppt,logram,guidelines,Drain}, we use two most commonly used metrics to evaluate the effectiveness of log parsers: Group Accuracy and Parsing Accuracy. %\peter{we need more descriptions/formula on how we compute PA and GA. }

\noindent{\underline{\textit{\textbf{Group Accuracy (GA):}}}}
Grouping Accuracy~\cite{tools} is a metric used in log parsing to evaluate the extent to which log messages belonging to the same template are correctly grouped together by a parser. 
GA is defined as the ratio of correctly grouped log messages to the total number of log messages. For a log message to be considered correctly grouped, it must be assigned to the same group as other log messages that share the same underlying template. 
%Specifically, GA is calculated as the ratio of log messages that are correctly grouped to the total number of log messages. A log message is considered correctly grouped if it shares the same group with other logs as defined by the same template. \peter{see if we can define and show some formula or examples}
%\zeyang{\begin{equation}
%GA = \frac{N_{\text{correctly grouped}}}{N_{\text{total}}}
%\end{equation}}
%GA focuses on evaluating the log parser's ability to accurately identify and cluster log messages that share the same template. 
High GA indicates that the parser can effectively discern patterns within the log data and group similar log messages together. This can be crucial for various downstream log analysis tasks such as anomaly detection~\cite{khan2023impact,10.1007/978-3-030-88494-9_16}. 
Despite its usefulness, GA has limitations. 
%GA can be high even when the parsed templates are flawed. This means that a high GA score could potentially mask errors in dynamic variable extraction and template identification within the logs, giving a false impression of overall parsing accuracy.
GA can remain high even if the parsed templates are flawed. Namely, a high GA score might obscure errors in dynamic variable extraction and template identification within the logs, leading to a misleading perception of overall parsing accuracy.

\noindent{\underline{\textit{\textbf{Parsing Accuracy (PA):}}}}
Parsing Accuracy (PA)~\cite{liu2022uniparser} complements GA and is calculated as the ratio of accurately parsed log messages to the total number of log messages. 
%To complement GA, Parsing Accuracy~\cite{liu2022uniparser} is a metric for evaluating the accuracy of log parsers in correctly extracting templates and variables from log messages. 
For a log message to be deemed correctly parsed, both extracted static text and dynamic variables must match exactly with those specified in the ground truth. 
PA is a stricter metric because it requires a comprehensive match of all log components, not just their correct grouping. This distinction is crucial, as GA primarily evaluates the correct clustering of logs, while PA ensures precise parsing accuracy at the individual log message level.
%This metric reis also essential for applications such as anomaly detection, where 
Precise log parsing of the variables can also significantly impact the effectiveness of downstream log-based analyses~\cite{10.1109/ICSE48619.2023.00078}.

%PA is calculated as the ratio of log messages that are parsed accurately to the total number of log messages. For a log message to be considered correctly parsed under this metric, every component—both static text and dynamic variables must match exactly with those specified in the ground truth. \peter{see if we can define and show some formula for examples}
%\zeyang{\begin{equation} PA = \frac{N_{\text{correctly parsed}}}{N_{\text{total}}}
%\end{equation}}
%PA is stricter than GA as it requires a comprehensive match, distinguishing it from metrics like GA, which primarily assess the correct clustering of logs rather than their individual parsing accuracy. This metric is essential for applications such as anomaly detection, where precise log parsing can significantly impact the effectiveness of downstream log-based analyses. 

\section{Evaluation}
\label{sec:evaluation}
In this section, we evaluate \logllm by answering four research questions (RQs). 

\subsection*{RQ1: What is the effectiveness of \logllm? }
\label{RQ1}

\phead{Motivation.} Accuracy is the most critical factor for evaluating the effectiveness of log parsers. High accuracy in log parsing aids downstream log analysis tasks~\cite{khan2023impact,10.1007/978-3-030-88494-9_16}. In this RQ, we study the effectiveness of \logllm.
\begin{table*}
\centering
\caption{A comparison of the grouping accuracy (GA) and parsing accuracy (PA) for the state-of-the-art parsers and \logllm. 
}\label{tab:RQ1}
\vspace{-1em}
\scalebox{0.78}{
\setlength{\tabcolsep}{5.3mm}{
\begin{tabular}{l|cc|cc|cc|cc?cc}\toprule
&\multicolumn{2}{c|}{\textbf{AEL}} &\multicolumn{2}{c|}{\textbf{Drain}} &\multicolumn{2}{c|}{\textbf{LILAC}}  &\multicolumn{2}{c?}{\textbf{\tbase}} &\multicolumn{2}{c}{\textbf{\logllm}}\\
\cline{2-11}
&GA &PA &GA &PA &GA &PA &GA &PA &GA &PA  \\
\hline
HDFS &0.9994 &0.6213 &0.9990 &0.6210 &\textbf{1.0000} &0.9480 &0.8295 &0.9663 &\textbf{1.0000} &\textbf{1.0000} \\
Hadoop &0.8230 &0.5350 &0.9210 &0.5410 &0.9240 &0.7850 &0.8376 &0.8370 &\textbf{0.9625} &\textbf{0.8706} \\
Spark &-- &-- &\textbf{0.8880} &0.3940 &0.8700 &0.7080 &0.2589 &0.4231 &0.8586 &\textbf{0.8887} \\
Zookeeper &\textbf{0.9960} &0.8420 &0.9940 &0.8430 &0.9970 &0.3780 &0.7192 &0.8408 &0.9932 &\textbf{0.8499} \\
BGL &0.9146 &0.4062 &\textbf{0.9190} &0.4070 &0.8335 &0.8239 &0.1439 &0.2440 &0.9024 &\textbf{0.9293} \\
HPC &0.7480 &0.7410 &0.7930 &0.7210 &\textbf{0.8450} &0.7350 &0.6423 &0.7070 &0.8440 &\textbf{0.9730} \\
Thunderbird &0.7859 &0.1635 &0.8310 &0.2160 &0.7940 &0.3860 &0.5790 &0.3472 &\textbf{0.8699} &\textbf{0.6940} \\
Linux &\textbf{0.9160} &0.0820 &0.6860 &0.1110 &0.7636 &0.7001 &0.4999 &0.8802 &0.9120 &\textbf{0.9017} \\
HealthApp &0.7250 &0.3110 &0.8620 &0.3120 &\textbf{0.9930} &0.6730 &0.8364 &0.9674 &0.8617 &\textbf{0.9735} \\
Apache &\textbf{1.0000} &0.7270 &\textbf{1.0000} &0.7270 &0.9970 &0.9920 &0.8680 &0.9900 &\textbf{1.0000} &\textbf{0.9960} \\
Proxifier &0.9740 &0.6770 &0.6920 &0.6880 &0.5060 &0.7780 &0.8276 &\textbf{0.9817} &0.5101 &0.8970 \\
OpenSSH &0.7050 &0.3640 &0.7070 &0.5860 &0.7480 &0.6550 &0.2183 &0.7812 &\textbf{0.8678} &0.4955 \\
OpenStack &0.7430 &0.0290 &0.7520 &0.0290 &0.5240 &0.4860 &\textbf{0.9517} &\textbf{0.9412} &0.8114 &0.8308 \\
Mac &0.7970 &0.2450 &0.7610 &0.3570 &0.8090 &0.4480 &0.6710 &0.6131 &\textbf{0.8141} &\textbf{0.6538} \\
\hline
Average &0.8559 &0.4418 &0.8432 &0.4681 &0.8289 &0.6783 &0.6345 &0.7514 &\textbf{0.8720} &\textbf{0.8538} \\
\bottomrule
\end{tabular}}
}
   \begin{tablenotes}
     \item \footnotesize{Note: The highest values of GA and PA for each system are highlighted in \textbf{bold}. The accuracy of AEL on the Spark dataset is excluded because it cannot complete parsing the whole dataset after running for 10 days. } 
   \end{tablenotes}
\vspace{-0.5em}
\end{table*}

\phead{Approach.} 
% We compare \logllm with other state-of-the-art log parsers, \peter{@zeyang, we need to say more about this, like AEL and Drain are traditional, LILAC and \tbase are LLM etc. (they show very good parsing results [cite])}including AEL, Drain, LILAC, and \tbase, which show high effectiveness or efficiency on small-scale datasets. 
We compare \logllm with other state-of-the-art log parsers, including AEL, Drain, LILAC, and \tbase.
AEL~\cite{AEL} and Drain~\cite{Drain} are two leading traditional syntax-based approaches that are efficient and perform better than most other syntax-based parsers~\cite{tools,loghub2}.
LILAC~\cite{lilac} and \tbase~\cite{LLMParser} are recently proposed LLM-based parsers with high parsing accuracy.
% We use LILAC~\cite{lilac} as one of our baselines. 
Since LILAC uses ChatGPT as the underlying LLM, for a fair comparison, we replace ChatGPT with the same open-source LLM (Llama3-8B~\cite{llama3}) that \logllm uses. We use T5-base~\cite{flan-t5} (240M parameters) as the LLM for LLMParser by following the prior work. Note that both LILAC and LLMParser require manually derived log templates as a few shot demonstrations. We follow the steps described in the papers to obtain these demonstrations. We evaluate the parsers using the LogHub-2.0 dataset and report both Grouping Accuracy (GA) and Parsing Accuracy (PA). %These metrics help in understanding how well a log parser groups similar logs together and how accurately it extracts log templates and variables. 

\phead{Results.} 
% \noindent{\bf {\em \logllm have a higher grouping accuracy and parsing accuracy compared to state-of-the-arts log parsers.}}
Table~\ref{tab:RQ1} shows the GA and PA for each log parser across different systems. \logllm achieved the highest GA and PA  values for most systems, indicating superior performance in both grouping and parsing logs. Across all systems, \logllm achieved an average GA of 0.8720 and an average PA of 0.8538, outperforming all other parsers. 

\noindent{\bf {\em 
\logllm shows superior GA and PA compared to the semi-supervised LLM-based parser -- LILAC. 
% Only a few demonstrations for in-context learning can not guide small parameter size model fit log parsing tasks on large-scale datasets. 
}}
Compared to \logllm, LILAC demonstrated lower performance with a GA of 0.8289 and PA of 0.6783. LILAC uses manually labeled logs as demonstrations for in-context learning to enhance parsing accuracy. However, when utilizing less powerful open-source LLMs with smaller parameter sizes (i.e., as opposed to ChatGPT), LILAC's performance declines significantly due to the limited ability of these models to capture complex log patterns with only a few demonstrations. 
Consequently, this can lead to inaccurate parsing of variables within the logs (a PA of 0.6783, while \logllm's PA is 0.8538). 
Unlike LILAC and \tbase, \logllm is an unsupervised log parser, eliminating the need for labeled logs to enhance the LLM's log parsing capabilities. The performance of \logllm is not dependent on the number of labeled logs, thus avoiding the limitations faced by semi-supervised approaches that require labeled logs for fine-tuning or in-context learning.

\noindent{\bf {\em Among all three LLM-based log parsers, \tbase shows the lowest GA, and the reason may be the limited number of fine-tuning samples that makes it hard to generalize to large-scale datasets.}} 
Among the three LLM-based log parsers (\tbase, LILAC, and \logllm), \tbase exhibited the lowest GA of 0.6345 and the second-highest PA of 0.7514. When parsing large-scale datasets, logs may exhibit many different variations, even if they share the same log template. Given that \tbase is fine-tuned using a small, labeled sample set from the target system, the limited number of log samples likely contributes to its inability to robustly identify logs with the same template across all instances and, thus, lower GA. This limitation becomes particularly evident in systems with more logs, such as BGL and Spark, where \tbase struggles to achieve high GA (0.1439 and 0.2589, respectively). Nevertheless, it is still able to identify all dynamic variables in a log with the second-highest PA among all five parsers, which shows the potentials of LLM-based parsers.

\noindent{\bf {\em Syntax-based log parsers generally have significantly lower PAs compared to LLM-based parsers, showing challenges in accurately identifying variables.}}
While AEL and Drain, as syntax-based parsers, show results similar to each other, they both exhibit lower GA compared to \logllm (1.84\% and 3.3\% lower, respectively) and significantly lower PA (48.25\% and 45.17\% lower, respectively). This performance disparity is likely linked to their heuristic-based nature, which relies on predefined rules to identify log features. While these rules can effectively classify logs with similar features, achieving reasonable GAs, their generic nature often fails to accurately recognize variables within different log templates, leading to poor PAs. In contrast, \logllm leverages pre-grouping and uses memory mechanisms to achieve high GA, and its LLM-based parsing process accurately identifies variables within grouped logs, resulting in superior PA.

%\rqboxc{\logllm achieves better GA and PA compared to state-of-the-arts. Although \logllm does not rely on labeled logs, \logllm shows better results in an unsupervised way than other LLM-based parsers. Compared to syntax-based approaches, \logllm significantly improves the PA.}

\rqboxc{\logllm achieves superior GA and PA compared to state-of-the-art parsers. Despite not relying on labeled logs, \logllm outperforms other LLM-based parsers that are semi-supervised. Additionally, \logllm significantly enhances PA compared to syntax-based approaches.}

\subsection*{RQ2: What is the efficiency of \logllm?}
\phead{Motivation.} Efficiency is crucial in log parsing since it 
directly impacts the practical usability of the parser in real-world applications. In this RQ, we study the parsers' efficiency. 
% Previous studies have shown that some log parsers show high efficiency in small-scale datasets, but when the log volume increases significantly, some log parsers may not be able to continue to parse efficiently.
%\noindent{\bf Approach.}
\begin{table*}
\centering
\caption{Number of logs and parsing time, in seconds, for the state-of-the-art (first four columns) and \logllm. 
}\label{tab:RQ2}
\vspace{-1em}
\scalebox{0.75}{
\setlength{\tabcolsep}{2.1mm}{
\begin{tabular}{l|r|r|r|r|r?rrrr}\toprule
% &\multicolumn{2}{c|}{\textbf{AEL}} &\multicolumn{2}{c|}{\textbf{Drain}} &\multicolumn{2}{c|}{\textbf{LILAC}} &\multicolumn{2}{c|}{\textbf{\tsmall}} &\multicolumn{2}{c?}{\textbf{\tbase}} &\multicolumn{2}{c}{\textbf{\logllm}}\\
% \cline{2-13}
% \begin{tabular}{lrrrrrrrrrrr}\toprule
& &\multicolumn{1}{c|}{\textbf{AEL}} &\multicolumn{1}{c|}{\textbf{Drain}}&\multicolumn{1}{c|}{\textbf{LILAC}} &\multicolumn{1}{c?}{\textbf{\tbase}} &\multicolumn{4}{c}{\textbf{\logllm}} \\
\cline{2-10}
&Log count &Total time &Total time &Total time &Total time &Total time &LLM query time &Grouping time &Memory search time \\
\hline
HDFS &11,167,740 &5,711.52 &1,343.56 &1,162.20 &148,097.72 &1,252.62 &273.67 &867.36 &111.59 \\
Hadoop &179,993 &361.54 &19.54 &4,747.52 &4,034.32 &285.76 &268.81 &11.95 &5.01 \\
Spark &16,075,117 & 10 days+&1,539.88 &3,346.08 &225,046.88 &1,752.40 &631.66 &764.12 &356.62 \\
Zookeeper &74,273 &3.22 &7.12 &1,702.46 &1,585.52 &52.23 &47.06 &4.75 &0.42 \\
BGL &4,631,261 &29,917.35 &501.09 &8,624.70 &90,526.27 &1,244.64 &857.82 &298.23 &88.59 \\
HPC &429,987 &18.00 &39.02 &388.88 &4,634.87 &539.76 &510.97 &27.45 &1.33 \\
Thunderbird &16,601,745 &25,199.44 &2,132.20 &16,316.03 &421,864.78 &8,659.29 &5,343.56 &1,466.77 &1,848.96 \\
Linux &23,921 &4.53 &2.61 &2,374.83 &1,031.82 &216.03 &213.42 &1.78 &0.82 \\
HealthApp &212,394 &976.74 &17.88 &1,182.27 &3,139.20 &103.33 &85.25 &10.38 &7.70 \\
Apache &51,977 &3.20 &5.42 &122.79 &1,056.53 &18.92 &15.19 &3.36 &0.37 \\
Proxifier &21,320 &1.69 &2.96 &681.65 &821.44 &871.52 &868.99 &2.47 &0.07 \\
OpenSSH &638,946 &1,338.67 &74.12 &1,134.35 &15,262.29 &89.37 &36.94 &49.00 &3.44 \\
OpenStack &207,632 &30.28 &60.18 &1,260.64 &7,558.24 &377.64 &330.66 &44.88 &2.09 \\
Mac &100,314 &10.79 &16.94 &15,930.81 &4,873.72 &5,935.77 &5,922.19 &9.26 &4.32 \\
\hline
Average &3,601,187 &4,890.54 &411.61 &4,212.52 &66,395.26 &1,528.52 &1,100.44 &254.41 &173.67 \\
\hline
Total &50,416,620 &17.66 hours &1.60 hours &16.38 hours &258.20 hours &5.94 hours &4.28 hours &0.99 hours &0.68 hours \\
\bottomrule
\end{tabular}}}
   % \begin{tablenotes}
   %   \item \footnotesize{Note: The running time of AEL on the Spark system is excluded because it cannot complete parsing the whole dataset in a reasonable time (10 days). } 
   % \end{tablenotes}
\vspace{-1em}
\end{table*}
\phead{Approach.} We measure the total parsing time required by \logllm and its individual components (i.e., LLM queries, grouping, and memory search), and the four baseline parsers to process logs from the LogHub-2.0 dataset. 
%We measure the total parsing time for every parser, and the time taken for individual components in 
%\logllm such as LLM queries, grouping, and memory search. 

\phead{Results.} \noindent{\bf {\em \logllm is 2.7 and 40 times faster than Lilac and \tbase, respectively.}} 
Table~\ref{tab:RQ2} shows the parsing time for each log parser across different systems. \logllm spends a total of 5.94 hours to parse logs from all 14 systems (50 million logs), which is significantly faster than other LLM-based parsers: LILAC (16 hours) and \tbase (258 hours). The parsing time for \logllm is mainly occupied by the LLM query time, which accounts for 72.05\% of the total processing time, followed by the grouping time, which constitutes 16.67\% of the overall duration.
%\noindent{\bf {\em \logllm is 2.7 and 40 times faster than Lilac and \tbase, respectively.}}
\tbase is the slowest among all LLM-based parsers because it %, requiring 40 times more time to parse all logs than \logllm. This inefficiency arises because \tbase
processes each log individually, and the vast quantity of logs linearly increases the number of model queries required. Even with a relatively lightweight model like T5-base, which has only 240 million parameters, querying to parse the logs individually is still slow and impractical for real-world applications. 
LILAC, with its cache design, eliminates the need to parse each log individually through an LLM, significantly speeding up the process compared to \tbase. However, LILAC still requires frequent model queries to update the templates in the cache, which limits its efficiency. In contrast, \logllm optimizes parsing times through its grouping and memory features, resulting in superior efficiency.

% \zeyang{compare with AEL and Drain. 
% AEL is efficient on small systems, but time increases dramatically on large systems.
% Drain is the fastest parser, but slower than \logllm's grouping time. Proving our grouping part is efficient.}
\noindent{\bf {\em AEL exhibits significant efficiency issues when parsing logs beyond certain sizes, while Drain maintains high efficiency across all datasets.}}
%AEL demonstrates a significant disparity in parsing time based on the volume of logs in a dataset. 
AEL can parse datasets with fewer than 100K logs within seconds but requires several hours or even days for datasets with over one million logs (e.g., we stopped AEL after running for 10 days when parsing the 16 million logs from Spark). 
This inefficiency is due to AEL's reliance on extensive comparisons between logs and identified templates, where the parsing time grows exponentially with respect to the number of logs and log templates. 
%As the volume of logs increases, not only do the comparisons between logs increase, but the number of identified templates also grows significantly, leading to an exponential increase in parsing time. 
In contrast, Drain, which uses a fixed-depth parsing tree, is the most efficient parser. % across various dataset sizes by employing a fixed-depth parsing tree. This approach has consistently placed Drain as the most efficient parser among those compared parsers. 
\logllm uses a grouping method similar to Drain’s, with a total grouping time amounting to 0.99 hours, which is less than Drain's total parsing time of 1.6 hours. This highlights the efficiency of \logllm's grouping process. While there is a slight slowdown due to the additional processing involved (5.94 hours compared to Drain's 1.6 hours), \logllm shows superior parsing effectiveness compared to Drain and is the second fastest log parser among the evaluates parsers. %parsing speed 

\rqboxc{\logllm enhances its efficiency by utilizing grouping and memory components, which reduces the number of LLM queries. \logllm demonstrates the highest efficiency across LLM-based parsers. }

\subsection*{RQ3: How does different settings impact the result of \logllm?}
% In this RQ, we explore how various settings and configurations affect the performance of \logllm. Understanding the impact of these settings is crucial for optimizing the parser's efficiency and accuracy in different operational contexts and for future research in the area. 

% \logllm has several components and settings that can be adjusted or replaced with others a. We evaluated these configurations, which included changes in sampling methods, the inclusion or exclusion of self-reflection processes, and variations in the number of log samples provided in the prompts.
\phead{Motivation.} \logllm implements multiple components to achieve effective and efficient log parsing. In this RQ, we explore how various settings and configurations affect the performance of \logllm. 
%Understanding the impact of these settings is crucial for optimizing the parser's efficiency and accuracy in different operational contexts and for future research in the area. 

\noindent{\bf Approach.} There are three general components in \logllm that can be adjusted or replaced: %includes has several components and settings that can be adjusted or replaced. We evaluated the impact of changing these configurations on both effectiveness and efficiency, which included changes in 
log selection from each group for prompting, the number of selected logs, and the inclusion or exclusion of self-reflection processes. 
To select diverse logs from the log group, we use Jaccard similarity to measure the similarity between every log pair. In this RQ, we also try random sampling and cosine similarity. 
Furthermore, we evaluate how changing the number of selected logs from 1 to 10 impacts the effectiveness. 
Finally, we compare the effect of removing the self-reflection component on the efficiency and effectiveness of \logllm.
% \peter{We need more details on how we change}\peter{and a short reminder on the similarity score - to select diverse logs. Something like: ``To select diverse logs from the same group ..., we use Jaccard similarity ...''}

\phead{Results.}
\noindent{\bf {\em Selecting representative logs based on Jaccard similarity outperforms using cosine similarity and random sampling.}} 
Table~\ref{tab:setting} shows the total time, GA, and PA of \logllm compared to replacing the log selection process with cosine similarity and random sampling. 
When employing cosine similarity to select representative logs, both GA and PA experienced declines of 2.5\% and 4.8\%, respectively, compared to using Jaccard similarity. This indicates that although cosine similarity is shown to be an effective similarity metric for text data~\cite{singhal2001modern}, it does not necessarily select logs that are representative enough for LLM to generalize log templates. However, we notice a slight reduction in execution time (3.3\%) when using cosine similarity. 
Similarly, using random sampling further reduces the processing time (by 8.2\%), but due to the lack of diversity in the sampled logs, both GA and PA are even lower, at 0.849 and 0.806, respectively. 

\begin{table}
\caption{\logllm performance under different settings. The numbers in the parenthesis indicate the percentage difference compared to the full version of \logllm. }\label{tab:setting}
% \scriptsize
\scalebox{0.79}{
\begin{tabular}{lcccc}\toprule
&\textbf{Total Time} &\textbf{GA} &\textbf{PA} \\
\midrule
\textbf{\logllm} &5.944 hours &0.872 &0.859 \\
\midrule
w/ cosine similarity  &5.745 (\textdownarrow 3.3\%) &0.85 (\textdownarrow 2.5\%) &0.818 (\textdownarrow 4.8\%) \\
w/ random sampling &5.458 (\textdownarrow 8.2\%) &0.849 (\textdownarrow 2.6\%) &0.806 (\textdownarrow 6.2\%) \\
\midrule
w/o self-reflection &3.292 (\textdownarrow 44.6\%) &0.81 (\textdownarrow 7.1\%) &0.777 (\textdownarrow 9.5\%) \\
\bottomrule
\end{tabular}}
\vspace{-0.1cm}
\end{table}
\noindent{\bf {\em Although the self-reflection mechanism requires additional processing time, it significantly enhances the parsing results of \logllm.}} 
Table~\ref{tab:setting} compares full version \logllm and \logllm without self-reflection in the total execution time, GA, and PA. 
Excluding the self-reflection component from \logllm results in a 44.6\% reduction in parsing time (from around six to three hours). However, removing self-reflection greatly decreases %this modification comes at a substantial cost, as 
both GA and PA %experience marked decreases of 
by 7.1\% and 9.5\%, respectively. This shows that self-reflection significantly enhances the parsing effectiveness of \logllm, although at the expense of increased overhead due to additional LLM queries. Therefore, in practical applications, the inclusion of the self-reflection component in \logllm can be determined based on the specific needs of effectiveness or efficiency.

\begin{figure}
\centering
\includegraphics[width=0.85\columnwidth]{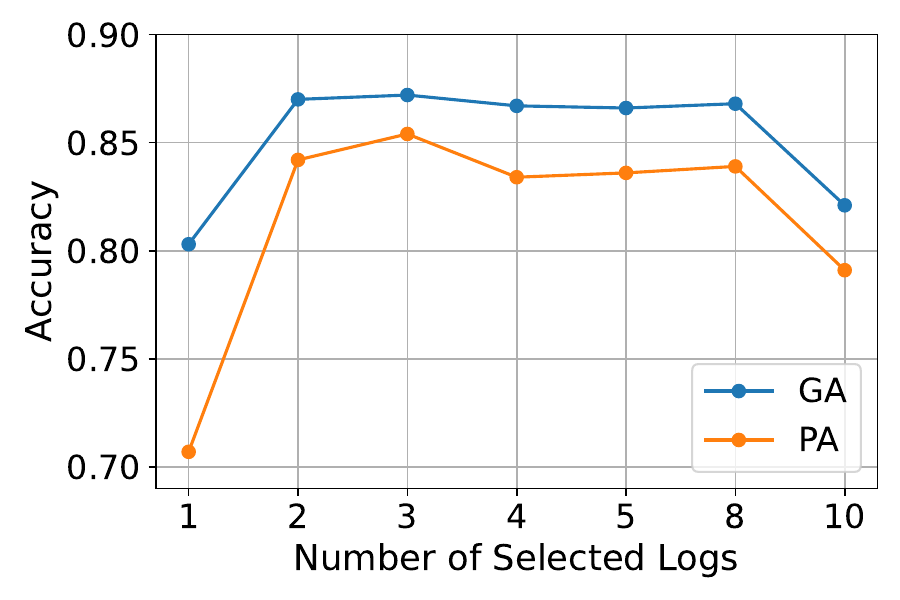}
\vspace{-1em}
\caption{GA and PA of \logllm using different numbers of selected logs in the prompt.}
\label{fig:sample_size}
\end{figure}

\noindent{\bf {\em During retrieval augmented log parsing, varying the number of selected logs affects the performance of  \logllm. Retrieving three logs into the prompt yields the highest effectiveness.}}
Fig~\ref{fig:sample_size} shows the \logllm performance with variations in the number of logs from a group retrieved into the prompts. 
\logllm maintains high accuracy across various sample sizes, with optimal performance achieved when the sample size is set to three, reaching peak values in both GA and PA. Notably, when the sample size is reduced to one, GA and PA drop to 0.80 and 0.70, respectively, representing a decline of 8.26\% and 18\% compared to a sample size of three. This reduction highlights the challenges LLM faces in parsing logs accurately without sufficient comparative data, such as multiple log comparisons or labeled logs. As the sample size increases from one to two, both GA and PA show significant improvements, peaking when the sample size reaches three. However, further increases in sample size from three to eight result in slight decreases in GA and PA, stabilizing around 0.865 and 0.835, respectively. This suggests that an excess of log samples may introduce noise, subsequently lowering performance~\cite{zhou2024can}. Importantly, when the sample size reaches 10, both GA and PA decrease compared to a sample size of eight. This decrease is attributed to prompt truncation caused by an overload of retrieved logs, which exceeds the context size of the LLM, resulting in incomplete input data.

\rqboxc{Using Jaccard similarity for log selection and LLM self-reflection enhances the parsing result, although they come with added overhead. Retrieving more logs within the prompt does not necessarily increase effectiveness; in fact, the optimal number of logs enables \logllm to reach peak accuracy is three.}

\subsection*{RQ4: What is the effectiveness of \logllm with different LLMs?}
\noindent{\bf Motivation.} 
Unlike previous parsers~\cite{chatgpt_howfar,lilac,divlog} that are based on commercial LLMs, \logllm employs open-source LLM to mitigate privacy concerns and monetary costs. Different LLMs exhibit varying capabilities due to their distinct architectures and pre-training data. In this RQ, we evaluate the performance of \logllm across various open-source LLMs.

\noindent{\bf Approach.}
%The baseline performance of \logllm is established using the default LLM (i.e., Llama3-8B). 
We selected three other open-source models (in addition to Llama3-8B) with similar parameter sizes to compare the log parsing performance with different LLMs, including Mistral-7B~\cite{jiang2023mistral7b}, CodeGemma-7B~\cite{codegemma}, and ChatGLM3-6B~\cite{glm2024chatglm}. These models are %representative and 
commonly used in research and practice. Mistral-7B shows strong text generation capabilities in small model sizes. CodeGemma-7B is pre-trained on code repositories and tailored for code-related tasks. ChatGLM3-6B is known for its bilingual conversational abilities. 

\phead{Results.}
Table~\ref{tab:RQ4} shows the parsing performance using various LLMs. %when replacing \logllm's default model, Llama3-8B, with three other open-source LLMs. 
Among all, %these LLMs, 
Llama3-8B achieved the best overall results.% across all models. 

\noindent{\bf {\em Compared to Llama3-8B, Mistral-7B requires a slightly longer parsing time, yet achieves a similar GA and a noticeable decline in PA. }}
Mistral-7B is a general model with training objectives and parameter sizes similar to Llama3-8B. However, it %shows comparable results in parsing time and GA but 
exhibits a lower PA, decreased by 14.6\%, with comparable results in parsing time and GA. This discrepancy in PA may be attributed to Llama3-8B's enhanced pre-training data, which includes more code~\cite{llama3}, and its larger parameter size. These factors likely contribute to Llama3-8B's superior ability to abstract variables within logs. %, a capability that Mistral-7B struggles to match.

\noindent{\bf {\em CodeGemma-7B has a better parsing speed, but both GA and PA face a decline compared to Llama3-8B.}}
CodeGemma-7B completes the parsing of all logs in only 71.5\% of the time required by Llama3-8B. %, demonstrating a significant improvement in processing speed.
However, CodeGemma-7B does not achieve comparably high accuracy, indicating that while it is capable of generating log templates that match the logs, it struggles to consistently and accurately abstract variables within these logs. Nevertheless, %Despite this reduction in effectiveness compared to Llama3-8B, 
using CodeGemma-7B still achieves higher GA and PA than other LLM-based parsers: LILAC and \tbase.

\noindent{\bf {\em As a conversational model, ChatGLM3-6B shows the worst result in parsing effectiveness and efficiency.}}
ChatGLM3-6B, pre-trained on a bilingual corpus in Chinese and English and optimized for conversations, %faces challenges when parsing semi-structured logs. Its pre-training goal 
does not include code in its pre-training data, which may have caused its bad parsing ability. This prevents ChatGLM3-6B from generating accurate log templates that can match the logs, necessitating increased model queries for self-reflection. Consequently, the parsing time for ChatGLM3-6B significantly increases by 145.1\% compared to LLaMA3. Despite undergoing extensive self-reflection, ChatGLM3-6B still fails to generate correct log templates. This leads to inferior results in both effectiveness and efficiency compared to other models, illustrating a clear disparity in performance when the pre-training background of the model does not match the specific task requirements.

\begin{table}
\centering
\caption{Parsing performance of \logllm using different LLMs. 
}\label{tab:RQ4}
%\vspace{-1em}

 \scalebox{0.77}{
\setlength{\tabcolsep}{3mm}{
\begin{tabular}{lcccc}\toprule
&\textbf{Total Time} &\textbf{GA} &\textbf{PA} \\\midrule
Llama3-8B &5.94 hours &0.872 &0.854 \\
Mistral-7B &6.78 (\textuparrow 14.1\%) &0.876 (\textuparrow 0.5\%) &0.729 (\textdownarrow 14.6\%) \\
CodeGemma-7B &4.25 (\textdownarrow 28.5\%) &0.814 (\textdownarrow 6.7\%) &0.752 (\textdownarrow 11.9\%) \\
ChatGLM3-6B &14.56 (\textuparrow 145.1\%) &0.837 (\textdownarrow 4\%) &0.600 (\textdownarrow 29.7\%) \\
\bottomrule
\end{tabular}
 }}
\vspace{-1em}
\end{table}

\rqboxc{Replacing the LLM leads to variations in effectiveness and performance. Among the four open-source models of similar sizes, Llama3-8B shows the best overall results.}

\section{Threats to Validity}
\label{sec:threat}
\noindent{\bf External validity.}
% \zeyang{Data leakage of LLM.
% Although we do not use labeled logs to fine-tune the LLM or use demonstrations in the prompt supervised. LLM may pre-train on open-source log data. 
% However, Fig~\ref{fig:sample_size} shows that only providing 1 log to LLM can not get good results. It proves that the low possibility of LLM having knowledge of the evaluated logs. LLM generates templates by analyzing the differences and commonalities from multiple retrieved logs.}
% \peter{does loghub2.0 included in the training? Can we check the timestamp?}
Data leakage is a potential risk of LLM-based log parsers~\cite{LLMParser,lilac}. Although \logllm does not involve using labeled logs for fine-tuning or in-context learning, there is a possibility that the LLM might have been pre-trained on publicly available log data. 
Our evaluation dataset with ground-truth templates was released on August 2023~\cite{loghub2} and Llama3-8B training knowledge cutoff from March 2023~\cite{llama3}, so the leakage risk should be minimal.
%Additionally, as shown in Figure~\ref{fig:sample_size}, the performance when provided with only a single log is lower than multiple logs, which proves that the template generation of LLM relies on analyzing the variations and similarities across multiple logs it retrieves instead of pre-training knowledge. Overall, data leakage in our experiments is not a significant issue.
The log format may also affect our result, but the datasets used are large and cover logs from various systems in different formats. Future studies are needed to evaluate \logllm on logs from other systems. 

\noindent{\bf Internal validity.}
% \zeyang{Adopting Llama3-8B as the base model. Future models may perform better. Llama3-8B is the latest released state-of-the-art small-size LLM and shows a promising result.}
\logllm employs Llama3-8B as its base model due to its promising results in many tasks and the relatively small size~\cite{llama3}. We also compared the results across various open-source LLMs and found differences. Future research is needed to evaluate LLM-based parsers' performance when more advanced LLMs are released in the future. The effectiveness of \logllm could be influenced by specific parameter settings (e.g., the number of logs selected for prompting). Our evaluations showed that these settings have an impact on the parsing results and discussed the optimal settings. Future studies are needed to evaluate the settings on other datasets. %Continued improvements in LLMs could provide better model support for \logllm to get a more effective and efficient log parsing practice in the future.
%It is anticipated that future LLMs may offer enhanced performance due to advancements in model architecture and training methodologies. 

\noindent{\bf Construct validity.}
% \zeyang{random output of LLM. temperature=0}
To mitigate the effects of randomness in evaluating \logllm, the generation temperature of the model is set to zero. This adjustment ensures that experiments conducted under the same conditions are repeatable and that the results are stable.

\section{Conclusion}
\label{sec:conclusion}

In this paper, we introduced \logllm, an unsupervised log parsing technique utilizing open-source LLMs to effectively address the limitations of existing LLM-based and syntax-based parsers.
\logllm first groups logs that share a syntactic similarity in the static text but vary in the dynamic variable, using a fixed-depth grouping tree.
It then parses logs in these groups with three components: i) retrieval augmented generation using similarity scoring: identifies diverse logs within each group based on Jaccard similarity, aiding the LLM in differentiating static text from dynamic variables; ii) self-reflection: iteratively queries LLMs to refine log templates and enhance parsing accuracy; and iii) log template memory: store parsed templates to minimize LLM queries, thereby boosting parsing efficiency.
% By leveraging the power of open-source LLMs, \logllm efficiently analyzes the variability within log groups and generates accurate log templates without any labeled logs for fine-tuning or in-context learning. \logllm incorporates a self-reflection step that iteratively refines these log templates if needed to improve the parsing accuracy. 
% Further enhancing its practicality, \logllm integrates a log template memory mechanism, which reduces the need for repeated LLM queries, optimizing the overall parsing efficiency. 
Our comprehensive evaluations on LogHub-2.0, a public large-scale log dataset, demonstrate that \logllm achieves an average GA of 0.8720 and an average PA of 0.8538, outperforming state-of-the-art parsers (i.e., ILIAC~\cite{lilac} and LLMParser~\cite{LLMParser}) by 5\% and 25\%, respectively. \logllm parses logs from all 14 systems (50 million logs) in a total of 5.94 hours, which is 2.75 and 40 times faster than other LLM-based parsers
This marks a substantial advancement over traditional semantic-based and LLM-based parsers in an unsupervised way, confirming the robustness and effectiveness of our approach. Additionally, \logllm addresses the privacy and cost concerns associated with commercial LLMs, making it a highly efficient and secure solution for practical log parsing needs.

 %\newpage
\balance
\bibliographystyle{plainnat}
\bibliography{ref}
\end{document}